\pdfoutput=1
\documentclass[journal = jcim]{achemso}
\usepackage{amsmath}
\usepackage[hidelinks]{hyperref}

\DeclareMathOperator{\Tr}{Tr}

\title{From Local Atomic Environments to Molecular Information Entropy}

\author{Alexander Croy}
\affiliation{Institute of Physical Chemistry, Friedrich Schiller University Jena, 07737 Jena, Germany}
\email{alexander.croy@uni-jena.de}

\SectionNumbersOn

\begin{document}
\begin{abstract}
The similarity of local atomic environments is an important concept in many machine-learning techniques which find applications in computational chemistry and material science. Here, we present and discuss a connection between the information entropy and the similarity matrix of a molecule. The resulting entropy can be used as a measure of the complexity of a molecule. Exemplarily, we introduce and evaluate two specific choices for defining the similarity: one is based on a SMILES representation of local substructures and the other is based on the SOAP kernel. By tuning the sensitivity of the latter, we can achieve a good agreement between the respective entropies. Finally, we consider the entropy of two molecules in a mixture. The gain of entropy due to the mixing can be used as a similarity measure of the molecules. We compare this measure to the average and the best-match kernel. The results indicate a connection between the different approaches and demonstrate the usefulness and broad applicability of the similarity-based entropy approach.
\end{abstract}

\maketitle

\section{Introduction}\label{sec:intro}

The concept of similarity is directly linked to the notion of the complexity of an object: Decomposing the object into different, distinguishable units, the number of those units provides a measure of its complexity \cite{Rashevsky_1955}. In this sense, the complexity, or information content, of a molecule can be defined, and over the past decades, different measures have been proposed, e.g.\ \cite{Bonchev_1977,Bertz_1981,Bonchev_1982,Boettcher_2016,Sabirov_2021}. Unfortunately, the different complexity measures are often hardly comparable.
In this article, we establish a connection between the information entropy (in the sense of Shannon \cite{Shannon_1948a,Shannon_1948b}) and the similarity matrix constructed from the local atomic environments of a molecule. The resulting expression is analogous to the von-Neumann entropy \cite{vonNeumann_1932} and can be used as a general framework for quantifying molecular complexity.

Similarity also plays an important role in many machine learning techniques, like kernel-ridge regression (KRR) or Gaussian process regression (GPR) \cite{Rasmussen_2006}. In combination with descriptors of local atomic environments\cite{Behler_2007,Bartok_2010}, those methods have been very successful in different areas of computational chemistry and material science \cite{Deringer_2021}. The atomic environments are usually defined in terms of all atoms within a certain distance of a reference atom. Suitable descriptors are then found, for example, from an expansion of the density of atoms in the environment into radial basis functions and spherical harmonics, i.e.\ using a smooth overlap of atomic positions (SOAP) \cite{Bartok_2013}. The kernel function entering KRR or GPR is calculated from such descriptors and can be interpreted as a measure of the similarity of two local atomic environments. Comparing all pairs of environments in this way leads to the similarity matrix of the specific molecule. On the other hand, for learning and predicting the global properties of molecules, one can introduce the similarity between different molecules \cite{Nikolova_2003}. The latter can also be constructed from the similarity matrix of the local atomic environments \cite{De_2016}.

In order to calculate the molecular information entropy, we present two specific choices for defining a similarity function of local atomic environments: One is based on a graph representation of the molecule, choosing substructures around a reference atom and comparing the resulting SMILES strings \cite{Weininger_1988,Weininger_1989}. The second approach facilitates the aforementioned SOAP similarity kernel and thus uses the positions and atomic numbers of the atoms. Both approaches are suitable for automatized computational studies, and we present results for a selection of molecules from the QM9 dataset \cite{Ruddigkeit_2012,Ramakrishnan_2014}. 

Finally, we investigate the information entropy for pairs of molecules which leads to the mixing entropy. The latter is the maximal gain of information entropy upon mixing two molecules \cite{Sabirov_2021}. Based on this observation, we propose and construct a new similarity measure of molecules and compare it to previously studied kernels \cite{De_2016}. Our results demonstrate the usefulness and broad applicability of the similarity-based entropy approach.

\section{Methods}\label{sec:methods}

\subsection{Information Entropy}
Typically, information entropy is considered in contexts involving some kind of `experiment' or `process', where each time one of the events $A_1, A_2, \ldots, A_n$ occurs at random \cite{Khinchin_1957}. Knowing the probabilities $p_1, p_2, \ldots, p_n$ for those events, one can characterize the amount of {\em uncertainty} about the outcomes by introducing the {\em Shannon entropy}\cite{Shannon_1948a,Shannon_1948b} according to
\begin{equation}\label{eq:entropy}
    H(p_1, p_2, \ldots, p_n) = - \sum_{i=1}^n p_i \log p_i \;.
\end{equation}
The logarithm can be taken with respect to any base, but it is usually assumed to be base two. Moreover, we take $p_i \log p_i=0$ if $p_i=0$. One readily sees, that the entropy vanishes if the probability for one event is one and the others are zero accordingly. This describes an experiment with no uncertainty, because the outcome would always be the same. On the other hand, one has maximum uncertainty and thus maximal entropy if all events have the same probability $p_i=1/n$. In this case, $H=\log n$.

In the context of molecules and graphs \cite{Rashevsky_1955,Mowshowitz_1968} a different point of view might be more suitable. If we decompose the molecule (or a graph) into $n$ different parts (e.g., its atoms or vertices) and assign each of them with one of the equivalence classes (e.g., atom types) $A_1, A_2, \ldots, A_m$ ($m\le n$), then we can construct a finite scheme by associating the probability $p_i=n_i/n$ to the respective class. The number of parts we found for each class is denoted by $n_i$, i.e., $\sum_{i=1}^m n_i = n$. The entropy given by Eq.\ \eqref{eq:entropy} can be viewed as a measure of the {\em complexity} of the object. If all parts belong to the same class ($n_1=n$), the complexity is zero. Conversely, if all parts belong to a different class ($n_i=1$ and $n=m$), then the complexity is maximal for that particular system.

\subsection{Information Entropy from Similarity}
To obtain a connection between information entropy of a molecule $\mathcal{M}$ and the similarities of its atoms, we start from a similarity function as follows,
\begin{equation}\label{eq:sim_func}
    S(k,l) = \left\{\begin{array}{cc}
        1, & \text{if atoms k and l are equivalent},\\
        0 & \text{otherwise}\;.
    \end{array}\right.
\end{equation}
If two atoms $k$ and $l$ in the molecule are chemically, or otherwise, equivalent, the function yields $1$, and otherwise it gives $0$. Using the similarity function for all pairs of atoms yields the {\em similarity matrix} $\mathbf{S}(\mathcal{M})$ of the molecule. This matrix is symmetric, positive semi-definite and its trace equals the number of atoms in the molecule.

By permuting rows and columns of a similarity matrix of size $n\times n$ arising from Eq.\ \eqref{eq:sim_func}, it can always be written as a direct sum of matrices of ones, i.e., $\mathbf{S} \simeq \mathbf{1}_{n_1} \oplus \mathbf{1}_{n_2} \oplus \cdots \oplus \mathbf{1}_{n_m}$. The similarity matrix becomes block-diagonal with $m$ blocks of size $n_i\times n_i$, respectively. Consequently, it has $m$ non-zero eigenvalues, namely $\{n_1, n_2, \ldots, n_m\}$ and $n-m$ eigenvalues which are zero\footnote{Each matrix of ones $\mathbf{1}_{n_i} $ has one non-zero eigenvalue equal to $n_i$ and $n_i-1$ eigenvalues which are zero.}. 
This suggests that we can obtain a finite scheme directly from the similarity matrix since the non-zero eigenvalues of $\mathbf{S}$ divided by $n$ yield the required probabilities $p_i=n_i/n$. Moreover, we can directly calculate the associated entropy 
\begin{equation}\label{eq:entropy_from_sim}
    H(\mathbf{S}) = - \sum_{i=1}^m \frac{n_i}{n} \log \frac{n_i}{n} 
    = - \Tr \left(\frac{1}{n}\mathbf{S} \log \frac{1}{n}\mathbf{S}\right)\;.
\end{equation}
Here, $\Tr$ denotes the trace of the matrix and $\log$ is the matrix logarithm. The last expression is analogous to the von-Neumann entropy \cite{vonNeumann_1932} which is not only used in quantum mechanics, but also in the context of complex networks \cite{De_Domenico_2016}. This analogy suggests that we can generalize the similarity function Eq.\ \eqref{eq:sim_func} to any positive semi-definite and symmetric function with a value range $0\le S \le 1$. The expression for the entropy remains unchanged.

We can also introduce the {\em linear entropy} by expanding the logarithm to the lowest order:
\begin{equation}\label{eq:linear_entropy}
    H(\mathbf{S}) \approx
    H_{\rm L}(\mathbf{S})
    = - \Tr \left[\frac{1}{n}\mathbf{S} \left( \frac{1}{n}\mathbf{S} - \mathbf{I}\right)\right]
    = 1 - \frac{1}{n^2}\Tr \left(\mathbf{S}^2\right)\;.
\end{equation}
It is given in terms of the average of the squared elements of the similarity matrix. If the latter only contains zeros and ones, the latter average is equal to the average of the elements themselves. While being approximate, this expression can be easier to calculate, especially for large similarity matrices.

Of course, the main question is how to find a suitable similarity function. From a chemical point of view one might have different options for specifying chemical equivalence \cite{Bertz_1981,Boettcher_2016} or one might use graph-theoretic concepts \cite{Mowshowitz_1968,Bonchev_1982}. In the following, two approaches are presented and discussed which are also motivated by applicability in computational settings.


\subsection{Substructure-SMILES Similarity}\label{sec:smiles_sim}
As a first approach to find the similarity of atomic environments, we use a graph representation of the molecule under consideration. For each atom of the molecule, we select a subgraph which involves all atoms which are connected to the reference atom by at most $N$ bonds. For example, $N=1$ would entail the atom itself and the bonded neighbors. Each subgraph is then converted to a (canonical) SMILES string \cite{Weininger_1988,Weininger_1989} with the reference atom as starting point. In practice, we use rdkit \cite{rdkit} to generate the environments and the SMILES strings. The similarity function is then simply defined via the comparison of the respective SMILES strings of the subgraphs, i.e.,
\begin{equation}\label{eq:sim_smiles}
    S_{\rm SMILES}(k,l) = \left\{\begin{array}{cc}
        1, & \text{if } {\rm SMILES}_k = {\rm SMILES}_l\;,\\
        0, & \text{otherwise}\;.
    \end{array}\right.
\end{equation}
Instead of exact string comparison, one can also utilize any other string or SMILES similarity method \cite{Ozturk_2016}.

\begin{figure}[tb]
    \centering
    \includegraphics[width=0.99\textwidth]{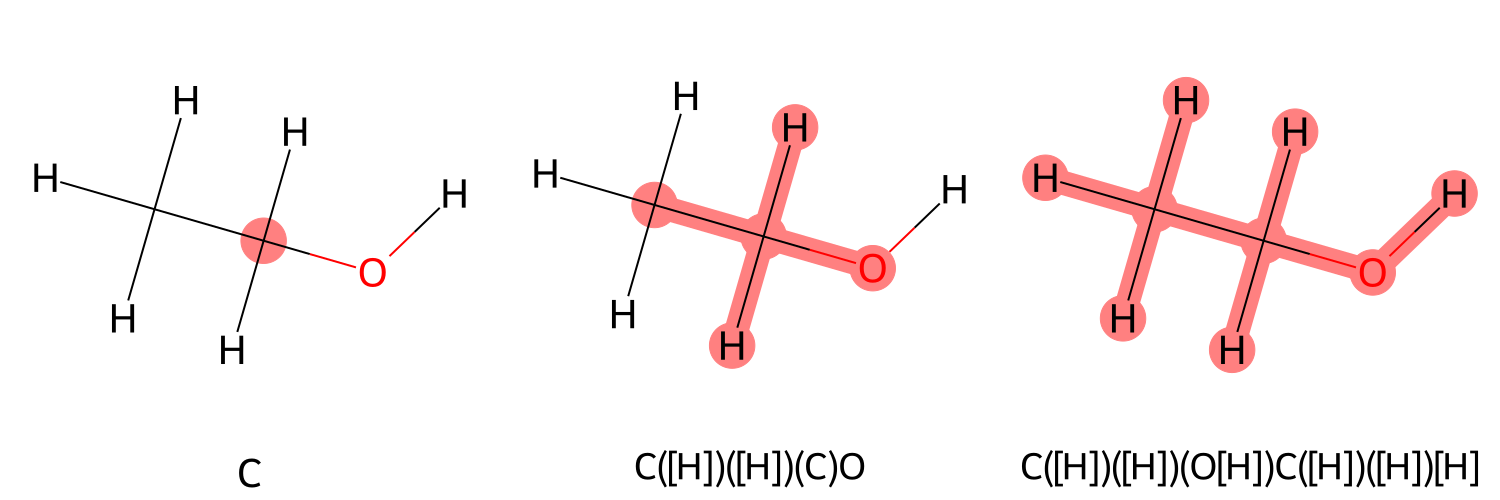}
    \includegraphics[width=0.32\textwidth]{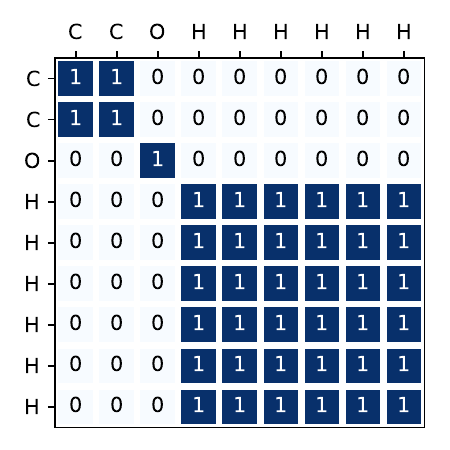}
    \includegraphics[width=0.32\textwidth]{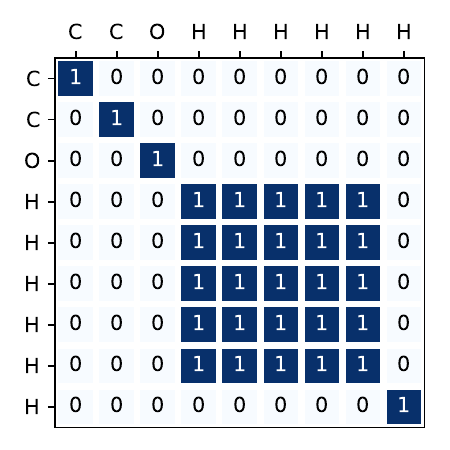}
    \includegraphics[width=0.32\textwidth]{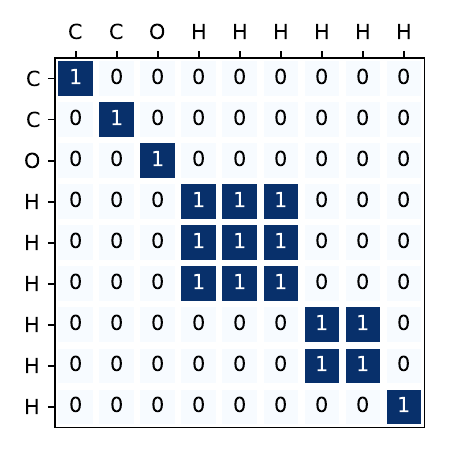}
    \caption{\label{fig:ethanol_smiles} Top row: Atomic environments for one of the carbon atoms in ethanol with $N=0,1,2$ (from left to right). The substructure SMILES strings are given below the structures. Bottom row: Similarity matrices obtained from Eq.\ \eqref{eq:sim_smiles} for $N=0,1,2$.}
\end{figure}

Figure \ref{fig:ethanol_smiles} illustrates the atomic environments of one of the carbon atoms in ethanol for different values of $N$. For $N=0$, the SMILES strings only contain the element symbol of the respective atom and therefore all hydrogen atoms, and all carbon atoms are found to be identical (lower right block in the similarity matrix). Already for $N=1$, this is no longer the case as the environments take neighboring atoms into account and the hydrogen atom bound to the oxygen is considered as distinct from the other hydrogens. The similarity matrices show the increasing differentiation between the different atoms until convergence is reached for $N\ge 2$ in this case.

\subsection{SOAP Similarity}
Another strategy for defining local atomic environments is the SOAP approach \cite{Bartok_2013,De_2016}. It is based on the representation of the local density of atoms $\rho_\mathcal{X}(\vec{r})$ in the vicinity of a central atom. The corresponding environment $\mathcal{X}$ is defined by a cutoff radius. Expanding the density in terms of radial basis functions $g_n(r)$ and spherical harmonics $Y_{lm}(\hat{r})$ leads to
\begin{equation}
    \rho^Z_\mathcal{X}(\vec{r}) = \sum_{n=1}^{n_{\rm max}}\sum_{l=0}^{l_{\rm max}}\sum_{m=-l}^{l} c^Z_{nlm} g_n(r) Y_{lm}(\hat{r})\;.
\end{equation}
Here, $r$ denotes the length and $\hat{r}$ the direction of the position vector $\vec{r}$, and $Z$ indicates the atomic species. 
The expansion coefficients $c^Z_{nlm}$ yield the rotationally invariant partial power spectrum,
\begin{equation}
    p^{Z Z'}_{n n' l}(\mathcal{X}) = \pi \sqrt{\frac{8}{2l+1}} \sum_{m=-l}^{l} \left(c_{nlm}^{Z}\right)^* c^{Z'}_{n' lm}\;.
\end{equation}
The elements $p^{Z Z'}_{n n' l}(\mathcal{X})$ can be collected into an unit length vector $\hat{p}(\mathcal{X})$ which is then used to define the similarity function for two atomic environments,
\begin{equation}\label{eq:sim_soap}
    S_{\rm SOAP}(k,l) = \left[\hat{p}(\mathcal{X}_k)\cdot \hat{p}(\mathcal{X}_l)\right]^\zeta \delta_{Z_k, Z_l}\;.
\end{equation}
It can be shown that $S_{\rm SOAP}$ is a positive definite function, which is obviously symmetric with respect to $k$ and $l$. The integer exponent $\zeta\ge 1$ can be used to increase the sensitivity of the similarity function \cite{Bartok_2013}. In the definition above, we have additionally enforced dissimilarity of environments centered around different species.

\section{Results}\label{sec:results}

\subsection{Molecular Information Entropies}
\begin{figure}[tb]
    \centering
    \includegraphics[width=0.99\textwidth]{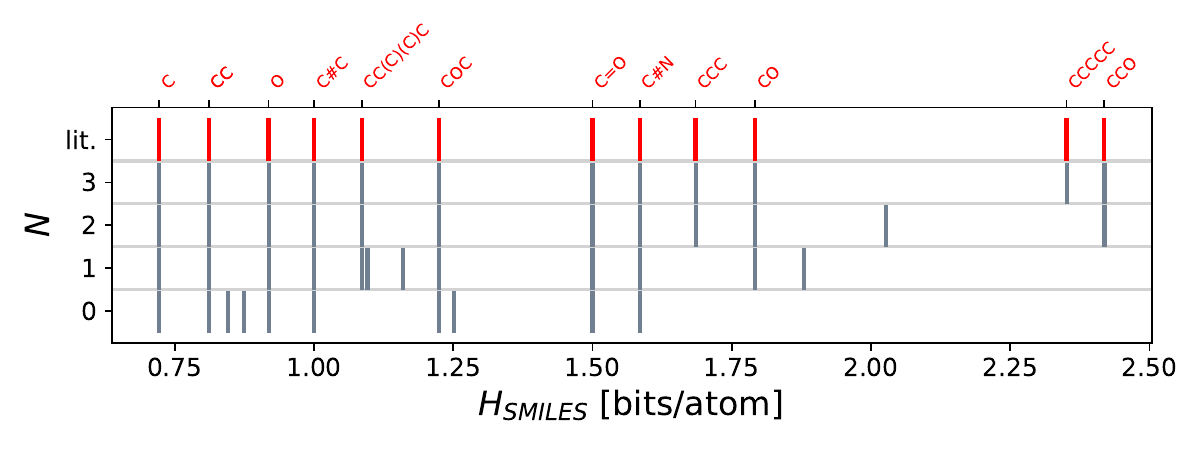}
    \caption{\label{fig:entropies_smiles} Molecular information entropies $H_{\rm SMILES}$ calculated from the substructure-SMILES similarity approach for the molecules denoted on the upper axis. The values are obtained for different sizes of the atomic environments ($N=0,1,2,3$) and compared to values from literature (topmost row) \cite{Sabirov_2018,Sabirov_2020}.} 
\end{figure}

First, we consider the molecular information entropies calculated from the substructure-SMILES similarity approach. To this end, we collected $13$ small molecules for which the information entropies are known based on symmetry and chemical intuition \cite{Sabirov_2018,Sabirov_2020}. Figure \ref{fig:entropies_smiles} 
shows a comparison of the entropies obtained from Eq.\ \eqref{eq:entropy_from_sim} for different sizes $N$ of the atomic environment. Consistent with the observation in Sec.\ \ref{sec:smiles_sim}, the entropy increases with increasing $N$ for some molecules until it converges to the expected value. In those cases, larger environments are necessary to achieve a sufficient differentiability between all the environments (as shown in Fig.\ \ref{fig:ethanol_smiles} for \textsf{CCO}).

\begin{figure}[tb]
    \centering
    \includegraphics[width=1.01\textwidth]{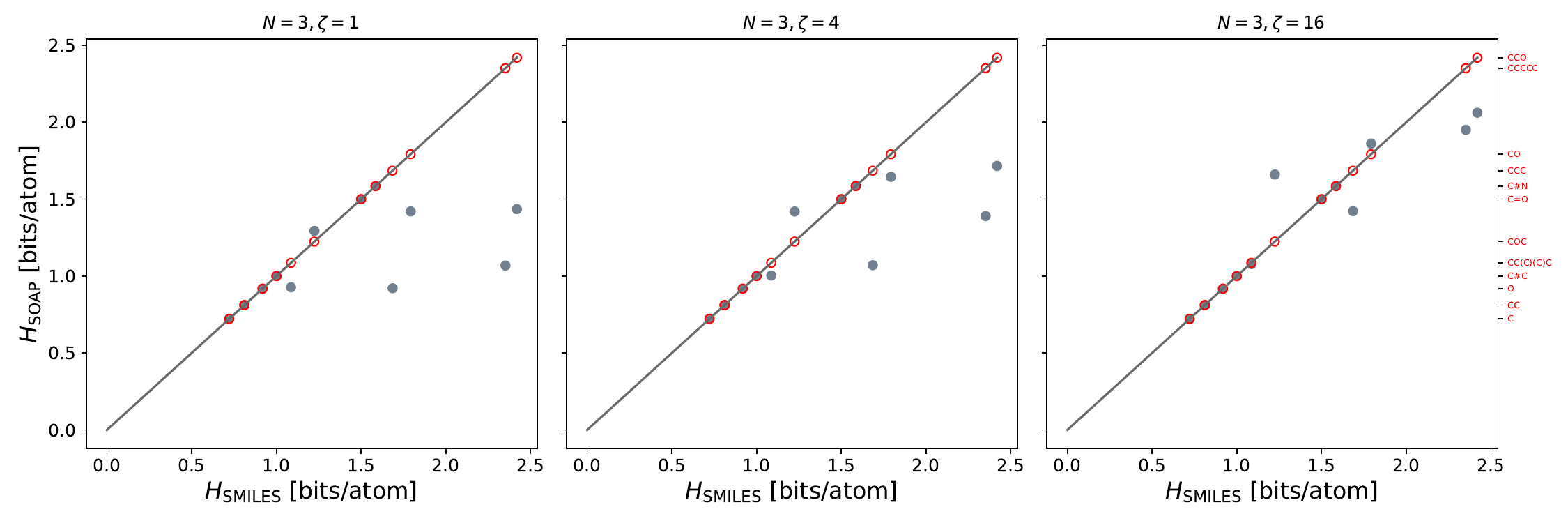}
    \caption{\label{fig:entropies_soap_smiles} Comparison of entropies obtained from the substructure-SMILES ($H_{\rm SMILES}$) and the SOAP approach ($H_{\rm SOAP}$), respectively. The red circles indicate values from literature \cite{Sabirov_2018,Sabirov_2020}. The different plots show SOAP entropies with increasing sensitivity exponent $\zeta$. Perfect matching is indicated by the straight lines.}
\end{figure}

In the next step, we use the ground state geometries of the $13$ molecules mentioned above from the QM9 dataset \cite{Ruddigkeit_2012,Ramakrishnan_2014} and calculate the SOAP descriptors for each atom using dscribe \cite{dscribe,dscribe2} (with $r_{\rm cut} = 6{\rm \AA}$, $n_{\rm max} = 10$, and $l_{\rm max} = 6$). 
The similarity matrix and entropy are then calculated via Eqs.\ \eqref{eq:sim_soap} and \eqref{eq:entropy_from_sim}, respectively, for a given sensitivity exponent $\zeta$. Figure \ref{fig:entropies_soap_smiles} shows a comparison of the corresponding entropies with the entropies found from the substructure-SMILES similarity approach (with $N=3$). As the sensitivity increases for larger $\zeta$, the entropies also increase since the similarity function further distinguishes the atomic environments. Apparently, the entropies do not converge in all cases to the SMILES entropies. The reason is that the functional form of the similarity function in Eq.\ \eqref{eq:sim_smiles} suppresses all entries which are $\le 1$ with increasing $N$ which leads in the extreme case to all environments being dissimilar and yields the maximum entropy.

\begin{figure}[tb]
    \centering
    \includegraphics[width=0.99\textwidth]{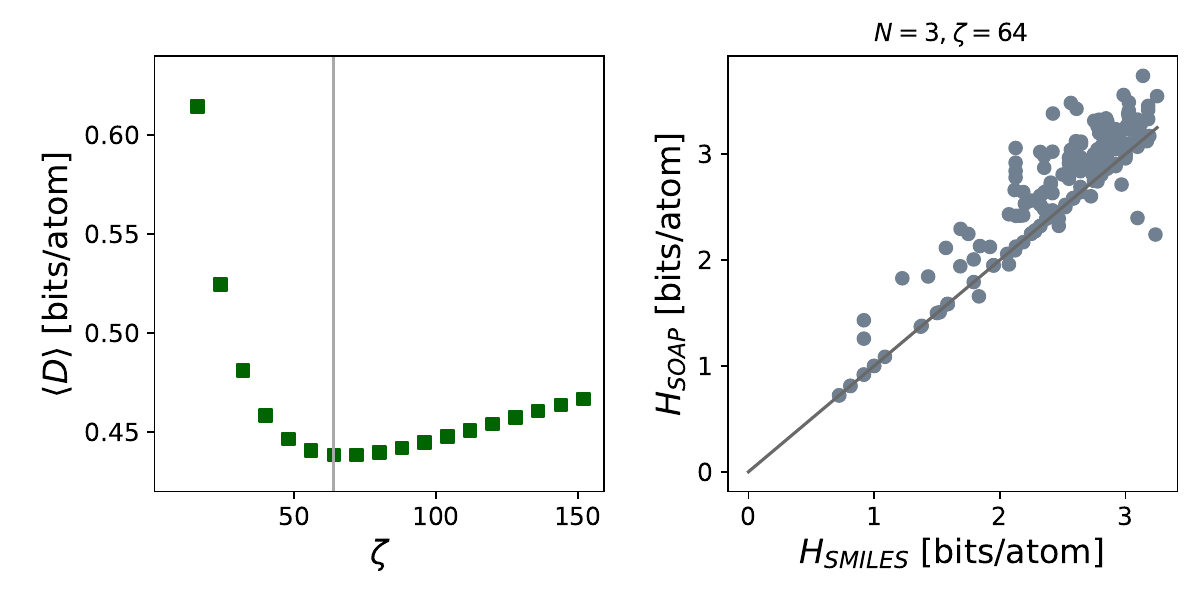}
    \caption{\label{fig:KL_divergence_soap_smiles} Left: Average Kullback-Leibler divergence $\langle D\rangle$ of the SMILES and SOAP similarity matrices of $174$ molecules. Right: Comparison of entropies obtained from the substructure-SMILES ($H_{\rm SMILES}$) and the SOAP approach ($H_{\rm SOAP}$) near the minimum of $\langle D\rangle$ for $174$ molecules. Perfect matching is indicated by the straight line.}
\end{figure}

To further characterize the similarities arising from the two approaches, we consider the substructure-SMILES similarity as references and compute the Kullback-Leibler (KL) divergence \cite{Kullback_1951} for the SOAP similarity of the same molecule,
\begin{equation}\label{eq:KL_divergence}
    D(\mathbf{S}_{\rm SMILES}(\mathcal{M}) \| \mathbf{S}_{\rm SOAP}(\mathcal{M}))
    = - \Tr \left(\frac{1}{n}\mathbf{S}_{\rm SMILES} \log \frac{1}{n}\mathbf{S}_{\rm SOAP}\right)
      - H_{\rm SMILES}(\mathcal{M})\;.
\end{equation}
The KL divergence is a positive function and becomes zero if the two similarity matrices are identical. As an illustration, we take the first $184$ molecules from the QM9 dataset [10 are excluded due to SMILES mismatch between rdkit and QM9] and compute the average KL divergence for different sensitivity exponents. From Fig.\ \ref{fig:KL_divergence_soap_smiles} one can see that the KL divergence initially decreases with increasing sensitivity - in accordance with our previous observations. Then, at $\zeta\approx 64$ it shows a minimum implying that the respective SOAP similarities on average match best to the SMILES similarities. For larger exponents, the KL divergence increases slowly as more and more entries in $\mathbf{S}_{\rm SOAP}$ are vanishing. Figure \ref{fig:KL_divergence_soap_smiles} also shows the comparison of the entropies obtained from the SOAP similarities for $\zeta=64$ and the SMILES-based similarities. We observe a good but not perfect agreement, which can again be attributed to the sweeping effect of the sensitivity exponent. It should be noted that the binary nature of the SMILES similarity is an extreme case which requires an atypical high sensitivity. 

\subsection{Mixing Entropy and Molecular Similarity}
One important question concerns the complexity or information entropy of mixtures of molecules as this directly applies to chemical reactions \cite{Karreman_1955,Sabirov_2018,Sabirov_2020}. We take two molecules, $\mathcal{M}_{\rm I}$ and $\mathcal{M}_{\rm II}$, and suppose that we have a similarity function $S$ which can be applied to any pair of atomic environments. The resulting similarity matrix $\mathbf{S}(\mathcal{M}_{\rm I}, \mathcal{M}_{\rm II})=\mathbf{S}_{\rm I,II}$ has a $2\times 2$ block structure with the diagonal blocks ($\mathbf{S}_{\rm I}$ and $\mathbf{S}_{\rm II}$) referring to the similarity between parts of each individual molecule, and the off-diagonal blocks ($\mathbf{S}_{\rm I,II}=\mathbf{S}_{\rm II,I}^T$) refer to similarity between environments belonging to different molecules.

If the two molecules are identical, all four blocks are equal to $\mathbf{S}_{\rm I}$ and the non-vanishing eigenvalues of  $\mathbf{S}_{\rm I,II}$ become those of $2 \mathbf{S}_{\rm I}$. Since the total number of parts is $2n_{\rm I}$, the entropy of the two molecules is the same as the one for the individual molecule. In the other case, if the two molecules do not share any equivalent atomic environments, the off-diagonal blocks are filled with zeros, and the similarity matrix can be written as a direct sum $\mathbf{S}_{\rm I+II}=\mathbf{S}_{\rm I}\oplus\mathbf{S}_{\rm II}$. In this case, the entropy becomes a weighted average of the individual entropies plus a term which can be called {\em entropy of mixing} \cite{Sabirov_2018,Sabirov_2020},
\begin{equation}
    H(\mathbf{S}_{\rm I+II})
    = H_{\rm mix}(n_{\rm I}, n_{\rm II})
    + \frac{n_{\rm I}}{n_{\rm I}+n_{\rm II}} H(\mathbf{S}_{\rm I})
    + \frac{n_{\rm II}}{n_{\rm I}+n_{\rm II}} H(\mathbf{S}_{\rm II})
\end{equation}
with
\begin{equation}\label{eq:twomol_ent_mix}
    H_{\rm mix}(n_{\rm I}, n_{\rm II})
    = - \frac{n_{\rm I}}{n_{\rm I}+n_{\rm II}} \log \frac{n_{\rm I}}{n_{\rm I}+n_{\rm II}} 
    - \frac{n_{\rm II}}{n_{\rm I}+n_{\rm II}} \log \frac{n_{\rm II}}{n_{\rm I}+n_{\rm II}}\;.
\end{equation}
For $n_{\rm I}=n_{\rm II}$, the mixing entropy is equal to $\log 2$ (or one bit per atom). For any pair of molecules, we can define the gain of entropy due to mixing via
\begin{equation}\label{eq:twomol_ent_gain}
    \Delta H(\mathcal{M}_{\rm I}, \mathcal{M}_{\rm II})
    = H(\mathbf{S}_{\rm I+II})
    - \frac{n_{\rm I}}{n_{\rm I}+n_{\rm II}} H(\mathbf{S}_{\rm I})
    - \frac{n_{\rm II}}{n_{\rm I}+n_{\rm II}} H(\mathbf{S}_{\rm II})\;.
\end{equation}
In general, this quantity will have values between $0$ and $H_{\rm mix}\le 1$.

\begin{figure}[tb]
    \centering
    \includegraphics[width=0.49\textwidth]{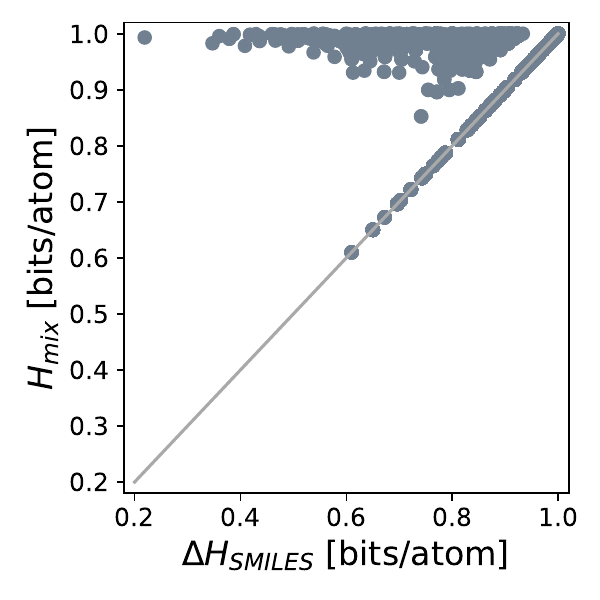}
    \caption{\label{fig:twomol_ent} Comparison of the gain of entropy due to mixing of two molecules $\Delta H_{\rm SMILES}$ and the respective mixing entropy $H_{\rm mix}$. The similarities are calculated using the substructure-SMILES approach (with $N=3$). Perfect matching is indicated by the straight line.}
\end{figure}

To illustrate the concept of mixing entropy, we take all pairs of the first $184$ molecules from the QM9 dataset and calculate the combined similarity matrices $\mathbf{S}_{\rm I,II}$ and the corresponding entropies $H(\mathbf{S}_{\rm I+II})$. In Fig.\ \ref{fig:twomol_ent} the latter are compared to the mixing entropies $H_{\rm mix}(n_{\rm I}, n_{\rm II})$ for the molecules using the substructure-SMILES approach to calculate the similarities (with $N=3$). For the data points on the diagonal, the molecules do not share equivalent atomic environments and $\Delta H = H_{\rm mix}$. However, there are a number of molecular pairs for which the entropy is smaller than the mixing entropy. 

The results for the mixing entropies suggest that one can use the ratio of the gain of entropy due to mixing $\Delta H(\mathcal{M}_{\rm I}, \mathcal{M}_{\rm II})$ and the mixing entropy $H_{\rm mix}(n_{\rm I}, n_{\rm II})$ as a measure for similarity of two molecules. This ratio is zero for two identical molecules and one for molecules not sharing atomic environments. Other approaches to compare two molecules based on similarities of local atomic environments include the average structural kernel and the best-match structural kernel \cite{De_2016}. The former is calculated by averaging the elements of the off-diagonal block in $\mathbf{S}_{\rm I,II}$, i.e.,
\begin{equation}
    \bar{K}^{(p)}(\mathcal{M}_{\rm I}, \mathcal{M}_{\rm II}) = \frac{1}{N_{\rm I} N_{\rm II}} \sum_{i\in \mathcal{M}_{\rm I}}\sum_{j\in \mathcal{M}_{\rm II}} S(i,j)^{p}\;.
\end{equation}
The best-match kernel can be formulated in terms of a rectangular assignment problem \cite{De_2016,Crouse_2016},
\begin{equation}
    \hat{K}^{(p)}(\mathcal{M}_{\rm I}, \mathcal{M}_{\rm II}) = \frac{1}{N_{\rm I}} \max_{\mathbf{X}}\sum_{i\in \mathcal{M}_{\rm I}}\sum_{j\in \mathcal{M}_{\rm II}} S(i,j)^{p} X_{i,j}\;.
\end{equation}
The matrix $\mathbf{X}$ contains only zeros and ones and additionally its elements are subject to $\sum_{j} X_{ij} = 1\, \forall i$ and  $\sum_{i} X_{ij} \le 1\, \forall j$. If $N_{\rm I} > N_{\rm II}$, the arguments $\mathcal{M}_{\rm I}$ and $\mathcal{M}_{\rm II}$ have to be interchanged. Here, we have generalized the two kernels by allowing arbitrary powers $p$ of the entries in $\mathbf{S}$ to be used while in Ref.\ \cite{De_2016} one is summing the elements of the similarity matrix ($p=1$) in both kernels. The reason for introducing $p$ is given by the expression of the linear entropy in Eq.\ \eqref{eq:linear_entropy}. There, only squared elements of the similarity matrix appear which suggests that taking the sum over the squared elements in $\bar{K}^{(p)}$ or $\hat{K}^{(p)}$, i.e.\ $p=2$, is more natural when comparing to the entropy-based similarity as also shown below.

\begin{figure}[tb]
    \centering
    \includegraphics[width=0.99\textwidth]{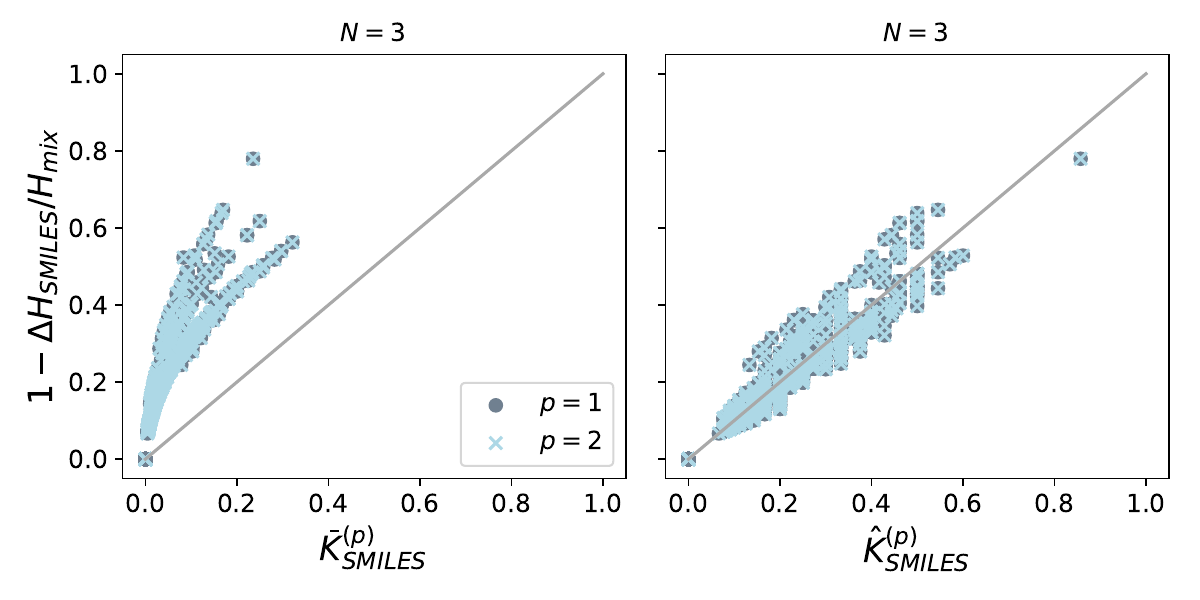}
    \includegraphics[width=0.99\textwidth]{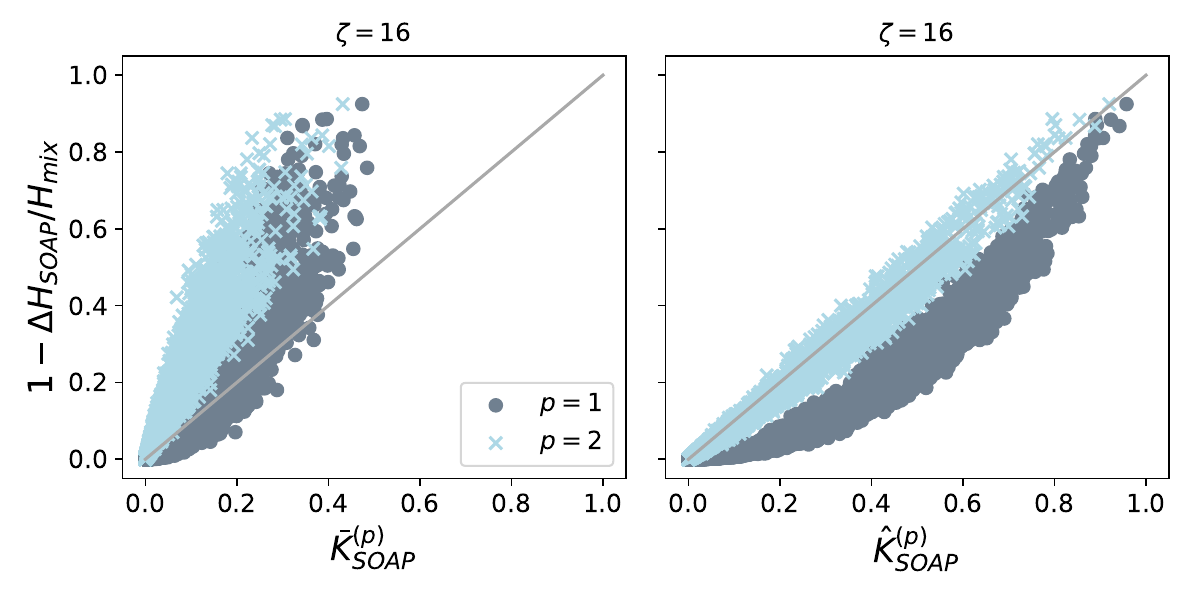}
    \caption{\label{fig:twomol_sims} Comparison of two molecular similarity kernels, $\bar{K}^{(p)}$ and $\hat{K}^{(p)}$, with a measure obtained via the gain of entropy due to mixing, $1-\Delta H/H_{\rm mix}$. Top row: Results for similarities calculated from the substructure-SMILES approach with environment size $N=3$. Bottom row: Results for similarities calculated from the SOAP approach with sensitivity exponent $\zeta=16$. Perfect matching is indicated by the straight lines.}
\end{figure}

Taking again the pairs of the $184$ molecules from before, we compute the different molecular similarities for the substructure-SMILES and the SOAP approaches and compare them in Fig.\ \ref{fig:twomol_sims}. One sees that for the SMILES approach, the results for $p=1$ and $p=2$ are identical, which is expected since $S_{ij}=S_{ij}^2$ in that case. On the other hand, for the SOAP approach one sees a clear dependence on $p$. The best-match kernel with $p=2$ yields similarities which agree on average with the entropy-based similarities. In contrast, the $p=1$ kernel gives rise to a systematic nonlinear deviation. With the average kernels one obtains similarities which are rather different from the entropy-based measure. In either case, it appears that the kernels with $p=2$ have a correspondence with the entropy-based similarities which is close to linear. The results show that the considered molecular similarity measures are related despite their different starting points. 

\section{Conclusions}\label{sec:conclusions}
In summary, we have presented and discussed a connection between the information entropy of a molecule and the similarity matrix of its local atomic environments. The entropy can be obtained from an expression, given by Eq.\ \eqref{eq:entropy_from_sim}, which is akin to the von-Neumann entropy. This relation provides a convenient framework for calculating the information entropy and also for analyzing its properties.

Two approaches for obtaining the similarity of atomic environments were given. The substructure-SMILES method is based on a graph representation of the molecule. The similarity is defined in terms of a comparison of SMILES strings for substructures. For a set of molecules, the entropies were shown to correspond to the known and expected values if the size of the substructures was sufficiently large. The second approach was based on SOAP descriptors which were obtained from positions and atomic numbers of atoms in a molecule. The sensitivity of the resulting similarities can be tuned by an integer exponent $\zeta$, and we found a good agreement between SOAP and SMILES based entropies for larger values of $\zeta$. It should be noted that any value $<1$ in the similarity matrix is suppressed with increasing $\zeta$, so that it is not expected that the values converge to the SMILES-based calculations. But the results show that both approaches can be used to estimate the information entropy of molecules. The choice of similarity function and the tuning of its hyperparameters (like $\zeta$) should be adapted to the given application.

Finally, we have investigated the entropy of a pair of molecules. For identical molecules, this entropy corresponds to the one of the individual molecule. If the molecules do not share any similar atomic environments, the ensemble entropy becomes equal to a weighted sum of the individual entropies plus the mixing entropy which only depends on the number of atoms in each molecule. In general, the entropy of the pair takes values between these two extremes. This was the motivation for defining the similarity of two molecules in terms of the entropy gain due to mixing. We compared this measure to two other similarity kernels (the average structural kernel and the best-match structural kernel) for $184$ molecules and found on average a very good agreement with a modified best-match kernel. Thus, the entropy-based molecular similarity provides an alternative measure for comparing molecules with a strong anchoring in the framework of molecular information entropies.

\section*{Data Availability}
The QM9 dataset is available at \url{https://doi.org/10.6084/m9.figshare.c.978904.v5}. The python code to calculate similarities and entropies can be found at \url{https://github.com/CoMeT4MatSci/molent}.

\begin{acknowledgement}
Inspiration and support is kindly acknowledged from the project `Olfactorial Perceptronics' (No.\ 9B396) funded by the Volkswagen foundation and SPP 2363 `Molecular Machine Learning' (No.\ GR4482/6) of the German Research Foundation. Stefanie Gr\"{a}fe is acknowledged for her input and helpful comments on the manuscript.
\end{acknowledgement}

\bibliography{references}

\providecommand{\latin}[1]{#1}
\makeatletter
\providecommand{\doi}
  {\begingroup\let\do\@makeother\dospecials
  \catcode`\{=1 \catcode`\}=2 \doi@aux}
\providecommand{\doi@aux}[1]{\endgroup\texttt{#1}}
\makeatother
\providecommand*\mcitethebibliography{\thebibliography}
\csname @ifundefined\endcsname{endmcitethebibliography}  {\let\endmcitethebibliography\endthebibliography}{}
\begin{mcitethebibliography}{33}
\providecommand*\natexlab[1]{#1}
\providecommand*\mciteSetBstSublistMode[1]{}
\providecommand*\mciteSetBstMaxWidthForm[2]{}
\providecommand*\mciteBstWouldAddEndPuncttrue
  {\def\EndOfBibitem{\unskip.}}
\providecommand*\mciteBstWouldAddEndPunctfalse
  {\let\EndOfBibitem\relax}
\providecommand*\mciteSetBstMidEndSepPunct[3]{}
\providecommand*\mciteSetBstSublistLabelBeginEnd[3]{}
\providecommand*\EndOfBibitem{}
\mciteSetBstSublistMode{f}
\mciteSetBstMaxWidthForm{subitem}{(\alph{mcitesubitemcount})}
\mciteSetBstSublistLabelBeginEnd
  {\mcitemaxwidthsubitemform\space}
  {\relax}
  {\relax}

\bibitem[Rashevsky(1955)]{Rashevsky_1955}
Rashevsky,~N. Life, information theory, and topology. \emph{The Bulletin of Mathematical Biophysics} \textbf{1955}, \emph{17}, 229--235\relax
\mciteBstWouldAddEndPuncttrue
\mciteSetBstMidEndSepPunct{\mcitedefaultmidpunct}
{\mcitedefaultendpunct}{\mcitedefaultseppunct}\relax
\EndOfBibitem
\bibitem[Bonchev and Trinajsti{\'c}(1977)Bonchev, and Trinajsti{\'c}]{Bonchev_1977}
Bonchev,~D.; Trinajsti{\'c},~N. Information theory, distance matrix, and molecular branching. \emph{The Journal of Chemical Physics} \textbf{1977}, \emph{67}, 4517--4533\relax
\mciteBstWouldAddEndPuncttrue
\mciteSetBstMidEndSepPunct{\mcitedefaultmidpunct}
{\mcitedefaultendpunct}{\mcitedefaultseppunct}\relax
\EndOfBibitem
\bibitem[Bertz(1981)]{Bertz_1981}
Bertz,~S.~H. The first general index of molecular complexity. \emph{Journal of the American Chemical Society} \textbf{1981}, \emph{103}, 3599--3601\relax
\mciteBstWouldAddEndPuncttrue
\mciteSetBstMidEndSepPunct{\mcitedefaultmidpunct}
{\mcitedefaultendpunct}{\mcitedefaultseppunct}\relax
\EndOfBibitem
\bibitem[Bonchev and Trinajsti{\'c}(1982)Bonchev, and Trinajsti{\'c}]{Bonchev_1982}
Bonchev,~D.; Trinajsti{\'c},~N. Chemical information theory: Structural aspects. \emph{International Journal of Quantum Chemistry} \textbf{1982}, \emph{22}, 463--480\relax
\mciteBstWouldAddEndPuncttrue
\mciteSetBstMidEndSepPunct{\mcitedefaultmidpunct}
{\mcitedefaultendpunct}{\mcitedefaultseppunct}\relax
\EndOfBibitem
\bibitem[B{\"o}ttcher(2016)]{Boettcher_2016}
B{\"o}ttcher,~T. An Additive Definition of Molecular Complexity. \emph{Journal of Chemical Information and Modeling} \textbf{2016}, \emph{56}, 462--470\relax
\mciteBstWouldAddEndPuncttrue
\mciteSetBstMidEndSepPunct{\mcitedefaultmidpunct}
{\mcitedefaultendpunct}{\mcitedefaultseppunct}\relax
\EndOfBibitem
\bibitem[Sabirov and Shepelevich(2021)Sabirov, and Shepelevich]{Sabirov_2021}
Sabirov,~D.~S.; Shepelevich,~I.~S. Information Entropy in Chemistry: An Overview. \emph{Entropy} \textbf{2021}, \emph{23}, 1240\relax
\mciteBstWouldAddEndPuncttrue
\mciteSetBstMidEndSepPunct{\mcitedefaultmidpunct}
{\mcitedefaultendpunct}{\mcitedefaultseppunct}\relax
\EndOfBibitem
\bibitem[Shannon(1948)]{Shannon_1948a}
Shannon,~C.~E. A mathematical theory of communication. \emph{The Bell System Technical Journal} \textbf{1948}, \emph{27}, 379--423\relax
\mciteBstWouldAddEndPuncttrue
\mciteSetBstMidEndSepPunct{\mcitedefaultmidpunct}
{\mcitedefaultendpunct}{\mcitedefaultseppunct}\relax
\EndOfBibitem
\bibitem[Shannon(1948)]{Shannon_1948b}
Shannon,~C.~E. A mathematical theory of communication. \emph{The Bell System Technical Journal} \textbf{1948}, \emph{27}, 623--656\relax
\mciteBstWouldAddEndPuncttrue
\mciteSetBstMidEndSepPunct{\mcitedefaultmidpunct}
{\mcitedefaultendpunct}{\mcitedefaultseppunct}\relax
\EndOfBibitem
\bibitem[von Neumann(1932)]{vonNeumann_1932}
von Neumann,~J. \emph{Mathematische Grundlagen der Quantenmechanik}; Springer Berlin, 1932\relax
\mciteBstWouldAddEndPuncttrue
\mciteSetBstMidEndSepPunct{\mcitedefaultmidpunct}
{\mcitedefaultendpunct}{\mcitedefaultseppunct}\relax
\EndOfBibitem
\bibitem[Rasmussen and Williams(2006)Rasmussen, and Williams]{Rasmussen_2006}
Rasmussen,~C.~E.; Williams,~C. K.~I. \emph{Gaussian Processes for Machine Learning}; The MIT Press, 2006\relax
\mciteBstWouldAddEndPuncttrue
\mciteSetBstMidEndSepPunct{\mcitedefaultmidpunct}
{\mcitedefaultendpunct}{\mcitedefaultseppunct}\relax
\EndOfBibitem
\bibitem[Behler and Parrinello(2007)Behler, and Parrinello]{Behler_2007}
Behler,~J.; Parrinello,~M. Generalized Neural-Network Representation of High-Dimensional Potential-Energy Surfaces. \emph{Phys. Rev. Lett.} \textbf{2007}, \emph{98}, 146401\relax
\mciteBstWouldAddEndPuncttrue
\mciteSetBstMidEndSepPunct{\mcitedefaultmidpunct}
{\mcitedefaultendpunct}{\mcitedefaultseppunct}\relax
\EndOfBibitem
\bibitem[Bart\'ok \latin{et~al.}(2010)Bart\'ok, Payne, Kondor, and Cs\'anyi]{Bartok_2010}
Bart\'ok,~A.~P.; Payne,~M.~C.; Kondor,~R.; Cs\'anyi,~G. Gaussian Approximation Potentials: The Accuracy of Quantum Mechanics, without the Electrons. \emph{Phys. Rev. Lett.} \textbf{2010}, \emph{104}, 136403\relax
\mciteBstWouldAddEndPuncttrue
\mciteSetBstMidEndSepPunct{\mcitedefaultmidpunct}
{\mcitedefaultendpunct}{\mcitedefaultseppunct}\relax
\EndOfBibitem
\bibitem[Deringer \latin{et~al.}(2021)Deringer, Bart{\'o}k, Bernstein, Wilkins, Ceriotti, and Cs{\'a}nyi]{Deringer_2021}
Deringer,~V.~L.; Bart{\'o}k,~A.~P.; Bernstein,~N.; Wilkins,~D.~M.; Ceriotti,~M.; Cs{\'a}nyi,~G. Gaussian Process Regression for Materials and Molecules. \emph{Chemical Reviews} \textbf{2021}, \emph{121}, 10073--10141\relax
\mciteBstWouldAddEndPuncttrue
\mciteSetBstMidEndSepPunct{\mcitedefaultmidpunct}
{\mcitedefaultendpunct}{\mcitedefaultseppunct}\relax
\EndOfBibitem
\bibitem[Bart{\'o}k \latin{et~al.}(2013)Bart{\'o}k, Kondor, and Cs{\'a}nyi]{Bartok_2013}
Bart{\'o}k,~A.~P.; Kondor,~R.; Cs{\'a}nyi,~G. On representing chemical environments. \emph{Physical Review B} \textbf{2013}, \emph{87}\relax
\mciteBstWouldAddEndPuncttrue
\mciteSetBstMidEndSepPunct{\mcitedefaultmidpunct}
{\mcitedefaultendpunct}{\mcitedefaultseppunct}\relax
\EndOfBibitem
\bibitem[Nikolova and Jaworska(2003)Nikolova, and Jaworska]{Nikolova_2003}
Nikolova,~N.; Jaworska,~J. Approaches to Measure Chemical Similarity -- a Review. \emph{QSAR \& Combinatorial Science} \textbf{2003}, \emph{22}, 1006--1026\relax
\mciteBstWouldAddEndPuncttrue
\mciteSetBstMidEndSepPunct{\mcitedefaultmidpunct}
{\mcitedefaultendpunct}{\mcitedefaultseppunct}\relax
\EndOfBibitem
\bibitem[De \latin{et~al.}(2016)De, Bart{\'o}k, Cs{\'a}nyi, and Ceriotti]{De_2016}
De,~S.; Bart{\'o}k,~A.~P.; Cs{\'a}nyi,~G.; Ceriotti,~M. Comparing molecules and solids across structural and alchemical space. \emph{Physical Chemistry Chemical Physics} \textbf{2016}, \emph{18}, 13754--13769\relax
\mciteBstWouldAddEndPuncttrue
\mciteSetBstMidEndSepPunct{\mcitedefaultmidpunct}
{\mcitedefaultendpunct}{\mcitedefaultseppunct}\relax
\EndOfBibitem
\bibitem[Weininger(1988)]{Weininger_1988}
Weininger,~D. SMILES, a chemical language and information system. 1. Introduction to methodology and encoding rules. \emph{Journal of Chemical Information and Computer Sciences} \textbf{1988}, \emph{28}, 31--36\relax
\mciteBstWouldAddEndPuncttrue
\mciteSetBstMidEndSepPunct{\mcitedefaultmidpunct}
{\mcitedefaultendpunct}{\mcitedefaultseppunct}\relax
\EndOfBibitem
\bibitem[Weininger \latin{et~al.}(1989)Weininger, Weininger, and Weininger]{Weininger_1989}
Weininger,~D.; Weininger,~A.; Weininger,~J.~L. SMILES. 2. Algorithm for generation of unique SMILES notation. \emph{Journal of Chemical Information and Computer Sciences} \textbf{1989}, \emph{29}, 97--101\relax
\mciteBstWouldAddEndPuncttrue
\mciteSetBstMidEndSepPunct{\mcitedefaultmidpunct}
{\mcitedefaultendpunct}{\mcitedefaultseppunct}\relax
\EndOfBibitem
\bibitem[Ruddigkeit \latin{et~al.}(2012)Ruddigkeit, van Deursen, Blum, and Reymond]{Ruddigkeit_2012}
Ruddigkeit,~L.; van Deursen,~R.; Blum,~L.~C.; Reymond,~J.-L. Enumeration of 166 Billion Organic Small Molecules in the Chemical Universe Database GDB-17. \emph{Journal of Chemical Information and Modeling} \textbf{2012}, \emph{52}, 2864--2875\relax
\mciteBstWouldAddEndPuncttrue
\mciteSetBstMidEndSepPunct{\mcitedefaultmidpunct}
{\mcitedefaultendpunct}{\mcitedefaultseppunct}\relax
\EndOfBibitem
\bibitem[Ramakrishnan \latin{et~al.}(2014)Ramakrishnan, Dral, Rupp, and von Lilienfeld]{Ramakrishnan_2014}
Ramakrishnan,~R.; Dral,~P.~O.; Rupp,~M.; von Lilienfeld,~O.~A. Quantum chemistry structures and properties of 134 kilo molecules. \emph{Scientific Data} \textbf{2014}, \emph{1}\relax
\mciteBstWouldAddEndPuncttrue
\mciteSetBstMidEndSepPunct{\mcitedefaultmidpunct}
{\mcitedefaultendpunct}{\mcitedefaultseppunct}\relax
\EndOfBibitem
\bibitem[Khinchin(1957)]{Khinchin_1957}
Khinchin,~A. \emph{Mathematical Foundations of Information Theory}; Dover Books on Mathematics; Dover Publications, 1957\relax
\mciteBstWouldAddEndPuncttrue
\mciteSetBstMidEndSepPunct{\mcitedefaultmidpunct}
{\mcitedefaultendpunct}{\mcitedefaultseppunct}\relax
\EndOfBibitem
\bibitem[Mowshowitz(1968)]{Mowshowitz_1968}
Mowshowitz,~A. Entropy and the complexity of graphs: I. An index of the relative complexity of a graph. \emph{The Bulletin of Mathematical Biophysics} \textbf{1968}, \emph{30}, 175--204\relax
\mciteBstWouldAddEndPuncttrue
\mciteSetBstMidEndSepPunct{\mcitedefaultmidpunct}
{\mcitedefaultendpunct}{\mcitedefaultseppunct}\relax
\EndOfBibitem
\bibitem[De~Domenico and Biamonte(2016)De~Domenico, and Biamonte]{De_Domenico_2016}
De~Domenico,~M.; Biamonte,~J. Spectral Entropies as Information-Theoretic Tools for Complex Network Comparison. \emph{Physical Review X} \textbf{2016}, \emph{6}\relax
\mciteBstWouldAddEndPuncttrue
\mciteSetBstMidEndSepPunct{\mcitedefaultmidpunct}
{\mcitedefaultendpunct}{\mcitedefaultseppunct}\relax
\EndOfBibitem
\bibitem[rdk()]{rdkit}
RDKit: Open-source cheminformatics. \url{https://www.rdkit.org}\relax
\mciteBstWouldAddEndPuncttrue
\mciteSetBstMidEndSepPunct{\mcitedefaultmidpunct}
{\mcitedefaultendpunct}{\mcitedefaultseppunct}\relax
\EndOfBibitem
\bibitem[{\"O}zt{\"u}rk \latin{et~al.}(2016){\"O}zt{\"u}rk, Ozkirimli, and {\"O}zg{\"u}r]{Ozturk_2016}
{\"O}zt{\"u}rk,~H.; Ozkirimli,~E.; {\"O}zg{\"u}r,~A. A comparative study of SMILES-based compound similarity functions for drug-target interaction prediction. \emph{BMC Bioinformatics} \textbf{2016}, \emph{17}\relax
\mciteBstWouldAddEndPuncttrue
\mciteSetBstMidEndSepPunct{\mcitedefaultmidpunct}
{\mcitedefaultendpunct}{\mcitedefaultseppunct}\relax
\EndOfBibitem
\bibitem[Sabirov(2018)]{Sabirov_2018}
Sabirov,~D.~S. Information entropy changes in chemical reactions. \emph{Computational and Theoretical Chemistry} \textbf{2018}, \emph{1123}, 169--179\relax
\mciteBstWouldAddEndPuncttrue
\mciteSetBstMidEndSepPunct{\mcitedefaultmidpunct}
{\mcitedefaultendpunct}{\mcitedefaultseppunct}\relax
\EndOfBibitem
\bibitem[Sabirov(2020)]{Sabirov_2020}
Sabirov,~D.~S. Information entropy of mixing molecules and its application to molecular ensembles and chemical reactions. \emph{Computational and Theoretical Chemistry} \textbf{2020}, \emph{1187}, 112933\relax
\mciteBstWouldAddEndPuncttrue
\mciteSetBstMidEndSepPunct{\mcitedefaultmidpunct}
{\mcitedefaultendpunct}{\mcitedefaultseppunct}\relax
\EndOfBibitem
\bibitem[Himanen \latin{et~al.}(2020)Himanen, J{\"a}ger, Morooka, Federici~Canova, Ranawat, Gao, Rinke, and Foster]{dscribe}
Himanen,~L.; J{\"a}ger,~M. O.~J.; Morooka,~E.~V.; Federici~Canova,~F.; Ranawat,~Y.~S.; Gao,~D.~Z.; Rinke,~P.; Foster,~A.~S. {DScribe: Library of descriptors for machine learning in materials science}. \emph{Computer Physics Communications} \textbf{2020}, \emph{247}, 106949\relax
\mciteBstWouldAddEndPuncttrue
\mciteSetBstMidEndSepPunct{\mcitedefaultmidpunct}
{\mcitedefaultendpunct}{\mcitedefaultseppunct}\relax
\EndOfBibitem
\bibitem[Laakso \latin{et~al.}(2023)Laakso, Himanen, Homm, Morooka, J{\"a}ger, Todorovi{\'c}, and Rinke]{dscribe2}
Laakso,~J.; Himanen,~L.; Homm,~H.; Morooka,~E.~V.; J{\"a}ger,~M.~O.; Todorovi{\'c},~M.; Rinke,~P. Updates to the DScribe library: New descriptors and derivatives. \emph{The Journal of Chemical Physics} \textbf{2023}, \emph{158}\relax
\mciteBstWouldAddEndPuncttrue
\mciteSetBstMidEndSepPunct{\mcitedefaultmidpunct}
{\mcitedefaultendpunct}{\mcitedefaultseppunct}\relax
\EndOfBibitem
\bibitem[Kullback and Leibler(1951)Kullback, and Leibler]{Kullback_1951}
Kullback,~S.; Leibler,~R.~A. On information and sufficiency. \emph{The annals of mathematical statistics} \textbf{1951}, \emph{22}, 79--86\relax
\mciteBstWouldAddEndPuncttrue
\mciteSetBstMidEndSepPunct{\mcitedefaultmidpunct}
{\mcitedefaultendpunct}{\mcitedefaultseppunct}\relax
\EndOfBibitem
\bibitem[Karreman(1955)]{Karreman_1955}
Karreman,~G. Topological information content and chemical reactions. \emph{The Bulletin of Mathematical Biophysics} \textbf{1955}, \emph{17}, 279--285\relax
\mciteBstWouldAddEndPuncttrue
\mciteSetBstMidEndSepPunct{\mcitedefaultmidpunct}
{\mcitedefaultendpunct}{\mcitedefaultseppunct}\relax
\EndOfBibitem
\bibitem[Crouse(2016)]{Crouse_2016}
Crouse,~D.~F. On implementing 2D rectangular assignment algorithms. \emph{IEEE Transactions on Aerospace and Electronic Systems} \textbf{2016}, \emph{52}, 1679--1696\relax
\mciteBstWouldAddEndPuncttrue
\mciteSetBstMidEndSepPunct{\mcitedefaultmidpunct}
{\mcitedefaultendpunct}{\mcitedefaultseppunct}\relax
\EndOfBibitem
\end{mcitethebibliography}

\end{document}